# Photonic orbital angular momentum with controllable orientation

Chenhao Wan [1,2], Jian Chen[1], Andy Chong [3,4], and Qiwen Zhan[1,*]

[1] School of Optical-Electrical and Computer Engineering, University of Shanghai for Science and Technology, Shanghai 200093, China

[2] School of Optical and Electronic Information, Huazhong University of Science and Technology, Wuhan, Hubei 430074, China

[3] Department of Electro-Optics and Photonics, University of Dayton, 300 College Park, Dayton, Ohio 45469, USA

[4] Department of Physics, University of Dayton, 300 College Park, Dayton, Ohio 45469, USA

*Corresponding author

E-mail: qwzhan@usst.edu.cn

Phone: 15685601103

Fax: 86-21-55272982

**Abstract**

Vortices are whirling disturbances commonly found in nature ranging from tremendously small scales in Bose-Einstein condensates to cosmologically colossal scales in spiral galaxies. An optical vortex, generally associated with a spiral phase, can carry orbital angular momentum (OAM). The optical OAM can either be in the longitudinal direction if the spiral phase twists in the spatial domain or in the transverse direction if the phase rotates in the spatiotemporal domain. In this article, we demonstrate the intersection of spatiotemporal vortices and spatial vortices in a wave packet. As a result of this intersection, the wave packet hosts a tilted OAM that provides an additional degree of freedom to the applications that harness the OAM of photons.

**Introduction**

Vortices, ubiquitous in nature, are circulating disturbances of liquid, gas or other media. They have been found in turbulent water, circulating air around wingtips, swirling galaxies and optics as well [1]. Optical vortices are generally associated with a spiral wavefront with phase singularities of zero intensity. The twisted wavefront gives rise to an azimuthal component of Poynting vector that contributes to an integrated orbital angular momentum (OAM) pointing along the beam axis. Each photon carries an OAM of $l\hbar$ where $\hbar$ is the reduced Planck's constant and $l$ is an integer, generally referred to as the topological charge [2]. The connection of vortex beams with optical OAM has spurred substantial theoretical and experimental research and found a wealth of applications in both classical and quantum optics [3-10].

Recent theoretical studies manifest that optical OAM does not have to be longitudinal but can be tilted to the optical axis [11,12]. The tilted OAM could be realized with a fast-moving observer close to the speed of light. Experimental progress has shown that a small fraction of optical energy can circulate in a spatiotemporal plane in a nonlinear interaction of an extremely high-power laser pulse and air [13]. Contrary to the longitudinal OAM that is associated with a spiral phase in the spatial domain, the transverse OAM roots in a spiral phase in the spatiotemporal domain that rotates around an axis perpendicular to the propagation direction. Although experimentally explored, it is still a challenging task to control and manipulate a spiral phase with circulating Poynting vector in a spatiotemporal plane in a linear manner. The difficulty has recently been overcome by forming a spiral phase in the spatial frequency - temporal frequency domain and retaining the spiral phase in the spatiotemporal domain through a two-dimensional spatiotemporal Fourier transform [14-16].

The intersection of spatial vortices has been reported in literature [17]. However, the interacting dynamics stays at the intersection point and does not travel with the beam. In this work, we experimentally demonstrate the intersection of spatiotemporal vortices and spatial vortices in an optical wave packet. The wave packet contains both screw and edge dislocations in phase. The intersection of two distinct types of optical vortices reveals interesting three-dimensional energy flow that travels at the speed of light. The combination of transverse OAM carried by spatiotemporal vortices and longitudinal OAM carried by spatial vortices gives rise to a tilted OAM with respect to the optical axis. The average three-dimensional OAM per photon remain unchanged after propagation in free space. The tilted OAM is fully controllable in value and orientation through the topological charges of the two types of vortices. Since orbital angular momentum plays an important role in light-matter interactions, the tilted OAM could be used for example to create optical spanners in arbitrary three-dimensional orientation with controllable torque. It can also be employed to provide an additional degree of freedom of optical OAM for applications such as optical tweezing, spin-orbit angular momentum coupling, and quantum communications. Furthermore, the platform to achieve tilted OAM could possibly be utilized as a laboratory tool to simulate relativistic effects as analyzed in [11].

**Results**

A wave packet is assumed to propagate in the $z$ direction with a vector potential $A$ polarized in the $x$ direction:

$$A = u(x, y, z')e^{ikz'}\mathbf{x}, \tag{1}$$

where $\mathbf{x}$ is the unit vector in the $x$ direction, $z$ is the propagation direction, $z'$ is the local frame coordinate given by $z' = z - ct$, $c$ is the speed of light in vacuum, and $k$ is the wave number. $u(x,y,z')$ is a complex scalar function describing the field distribution under paraxial approximation. A spatiotemporal vortex of topological charge $l_1$ and a spatial vortex of topological charge $l_2$ are both embedded within the wave packet. The complex scalar function $u$ is therefore given by,

$$u(x, y, z') = \left(\frac{x}{w_x} \pm i\,\text{sgn}(l_1)\frac{y}{w_y}\right)^{|l_1|}\left(\frac{x}{w_x} \pm i\,\text{sgn}(l_2)\frac{z'}{w_{z'}}\right)^{|l_2|}$$
$$\exp\left\{-\left[\left(\frac{x}{w_x}\right)^2 + \left(\frac{y}{w_y}\right)^2 + \left(\frac{z'}{w_{z'}}\right)^2\right]\right\} \tag{2}$$

where $w_x$, $w_y$ and $w_{z'}$ are the pulse dimension along the $x$, $y$, and $z'$ directions, respectively. Here the intensity distributions are assumed to be Gaussian in all three dimensions. The linear momentum density $\mathbf{g}$ is the time average of the real part of $\varepsilon_0(\mathbf{E} \times \mathbf{B})$ where $\mathbf{E}$ is the electric field, $\mathbf{B}$ is the magnetic field, and $\varepsilon_0$ is the permittivity in vacuum. Following the Lorentz gauge, the linear momentum density $\mathbf{g}$ is expressed in terms of the complex scalar function $u$ [18]:

$$\mathbf{g} = \frac{\varepsilon_0}{2}\left(\mathbf{E}^* \times \mathbf{B} + \mathbf{E} \times \mathbf{B}^*\right) = \frac{\varepsilon_0}{2}\left(c\frac{\partial u^*}{\partial x}\frac{\partial u}{\partial z'} + c\frac{\partial u}{\partial x}\frac{\partial u^*}{\partial z'} + i\omega u\frac{\partial u^*}{\partial x} - i\omega u^*\frac{\partial u}{\partial x}\right)\mathbf{x}$$
$$+ \frac{\varepsilon_0}{2}\left(-i\omega u^*\frac{\partial u}{\partial y} + i\omega u\frac{\partial u^*}{\partial y}\right)\mathbf{y} + \frac{\varepsilon_0}{2}\left(-i\omega u^*\frac{\partial u}{\partial z'} + i\omega u\frac{\partial u^*}{\partial z'} + 2\omega k|u|^2\right)\mathbf{z'} \tag{3}$$

where * denotes complex conjugate, $\mathbf{y}$ is the unit vector in the $y$ direction, and $\mathbf{z'}$ is the unit vector in the $z'$ direction. The angular momentum density is given by the cross product of the position vector $\mathbf{r}$ with the linear momentum density $\mathbf{g}$:

$$\mathbf{r} \times \mathbf{g} = \frac{\varepsilon_0}{2}\left[y\left(-i\omega u^*\frac{\partial u}{\partial z'} + i\omega u\frac{\partial u^*}{\partial z'} + 2\omega k|u|^2\right) - z'\left(-i\omega u^*\frac{\partial u}{\partial y} + i\omega u\frac{\partial u^*}{\partial y}\right)\right]\mathbf{x}$$
$$+ \frac{\varepsilon_0}{2}\left[z'\left(c\frac{\partial u^*}{\partial x}\frac{\partial u}{\partial z'} + c\frac{\partial u}{\partial x}\frac{\partial u^*}{\partial z'} + i\omega u\frac{\partial u^*}{\partial x} - i\omega u^*\frac{\partial u}{\partial x}\right) - x\left(-i\omega u^*\frac{\partial u}{\partial z'} + i\omega u\frac{\partial u^*}{\partial z'} + 2\omega k|u|^2\right)\right]\mathbf{y} \tag{4}$$
$$+ \frac{\varepsilon_0}{2}\left[x\left(-i\omega u^*\frac{\partial u}{\partial y} + i\omega u\frac{\partial u^*}{\partial y}\right) - y\left(c\frac{\partial u^*}{\partial x}\frac{\partial u}{\partial z'} + c\frac{\partial u}{\partial x}\frac{\partial u^*}{\partial z'} + i\omega u\frac{\partial u^*}{\partial x} - i\omega u^*\frac{\partial u}{\partial x}\right)\right]\mathbf{z'}$$

The average OAM per photon within the wave packet can be derived through a volume integral:

$$OAM/photon = \frac{\int_{-\infty}^{\infty} \mathbf{r} \times \mathbf{g} \, dV}{\varepsilon_0 \omega^2 \int_{-\infty}^{\infty} |u|^2 \, dV} h\upsilon. \tag{5}$$

The average OAM per photon is a vector that has three components pointing to the *x*, *y*, and *z'* directions, respectively. Based on eqs. (2) to (5), the *x*, *y*, and *z'* components of the OAM per photon are derived analytically. For example, if $l_1 = 3$ and $l_2 = 2$, the *x*, *y*, and *z'* components of the OAM per photon are calculated to be 0, $\frac{2(34w_x^2 + 7w_{z'}^2)}{41w_x w_{z'}}\hbar$, and $\frac{3(65w_x^2 + 17w_y^2)}{82w_x w_y}\hbar$, respectively. Therefore, the OAM per photon within the wave packet is $\frac{1}{82}\sqrt{27506 + w_x^2\left(\frac{38025}{w_y^2} + \frac{18496}{w_{z'}^2}\right) + \frac{2601w_y^2 + 784w_{z'}^2}{w_x^2}}\hbar$ pointing to the direction of $\theta$ degrees with respect to the -*y* axis in the *y-z'* plane, where $\theta = \tan^{-1}\left(\frac{195w_x^2 w_{z'} + 51w_y^2 w_{z'}}{136w_x^2 w_y + 28w_y w_{z'}^2}\right)$. Particularly, if $w_x$, $w_y$ and $w_{z'}$ are of equal length, the *x*, *y*, and *z'* components of the OAM are 0, $2\hbar$, and $3\hbar$ as expected. The theoretical analysis clarifies the fact that an average three-dimensionally oriented OAM can indeed be obtained in a nonrelativistic way through the intersection of a spatiotemporal vortex and a spatial vortex within one wave packet. The associated topological charges as well as the pulse dimensions play an important role in tuning the value and three-dimensional orientation of the tilted OAM.

Figure 1a shows simulated three-dimensional iso-intensity profiles of a wave packet with intersected spatiotemporal and spatial vortices. The intensity profiles are assumed to be Gaussian of equal length in all three directions. The topological charges of both vortices are assumed to be 1 (Eq. (2)). The spatiotemporal vortex twists in the *x-t* plane and forms a tunnel in the *y* direction (dark strip in Fig. 1b); the spatial vortex rotates in the *x-y* plane and forms a tunnel in the *t* direction (dark strip in Fig. 1c). After free-space propagation, the diffraction effect causes an interesting twist of the two tunnels, which can be understood as the consequence of the intersection of perpendicular vortices, at the intersection. A spatial vortex tunnel is twisted clockwise in the *x-t* plane by a spatiotemporal vortex of topological charge 1 (Fig. 1d). A spatiotemporal vortex tunnel is twisted counterclockwise in the *x-y* plane by a spatial vortex of topological 1 (Fig. 1e). Figures 1f and 1g show different twisting directions because the topological charges of both vortices are changed to -1.

Figures 2a to 2f show the experimental data of a generated wave packet that hosts the intersection of a spatiotemporal vortex of topological charge 1 and a spatial vortex of topological charge 1. The details of the experimental setup can be found in the supplementary information. Figures 2a to 2d display four typical slices of interference images in the *x-y* plane of the wave packet with the reference pulse. The positions of these slices are marked in Fig. 2e. The fringe patterns in Fig. 2a are dominantly influenced by the spatial vortex because the slicing position is far from the core of the spatiotemporal vortex. The most noticeable feature is a forklike pattern pointing downwards at the center that represents a spatial vortex of topological charge 1. As the reference pulse approaches the intersection center, the effect of the spatiotemporal vortex becomes more obvious. In Fig. 2b, the upper half of peripheral fringes start to bend to the right indicating an increasing phase shift between the upper and lower part of the wave packet. Bending fringes is a salient feature of a spatiotemporal vortex. Continued in Fig. 2c, the upper and lower peripheral fringes have a π phase difference. The dark fringes in the upper half are aligned with the bright fringes in the lower half. In Fig. 2d, the bending direction of peripheral fringes is switched. Slicing through the chirped wave packet from head to tail, the upper and lower peripheral fringes complete a 0 to π and π to 0 phase shift. The peripheral areas of fringe patterns in Figs. 2b to 2d have smaller spatial spiral phase gradient than the central areas. Therefore, the peripheral fringes mainly show the feature of a spatiotemporal vortex of topological charge 1. The central area of Fig. 2b to 2d is the intersection of the spatiotemporal vortex and the spatial vortex. The interaction of the two vortices is most intense in the central region and the features of both vortices are combined.

Extracting the optical field information from slices of fringes, the three-dimensional intensity distribution of the wave packet can be reconstructed [19-21]. Figure 2e and 2f show the wave packet from different views. In Fig. 2e, a vortex core appearing as a tunnel resulted from spatiotemporal phase singularity is clearly displayed. The spatiotemporal vortex tunnel penetrates through the wave packet along the *y* direction. In Fig. 2f, the spatial phase singularities result in a spatial vortex tunnel that penetrates all the way from head to tail of the wave packet. The spatiotemporal vortex tunnel is twisted clockwise at the core of the intersection of the two vortices. The spatial vortex tunnel, not surprisingly, is also twisted by the spatiotemporal vortex. Based on Eq. (5), the average OAM per photon of the wave packet is calculated to be $2.5\hbar$ pointing to the direction of 23.8 degrees with respect to the *-y* axis in the *y-z'* plane. Numerical simulation results show that the OAM of the wave packet is conserved during free space propagation.

Figures 2g to 2l shown the intersection of a spatiotemporal vortex of topological charge -1 and a spatial vortex of topological charge -1. Fig. 2g shows an interference pattern that is sliced far away from the intersection center. The spatial vortex feature is overwhelmed and a forklike pattern pointing upwards is shown at the center of the image. The change in the pointing direction of the forklike pattern is due to the fact that the topological charge of the spatial vortex is changed to -1. From Fig. 2h to 2j, the peripheral fringes display the features of a spatiotemporal vortex. Since the sign of topological charge of the spatiotemporal vortex is reversed, the upper half of peripheral fringes first bends to the left as shown in Fig. 2h until a π phase shift is accomplished as shown in Fig. 2i, and then bends to the right as shown in Fig. 2j. Figures 2k and 2l show the three-dimensional intensity reconstruction of the wave packet. The spatiotemporal vortex tunnel and the spatial vortex tunnel are both twisted at the intersection. The twisting direction, however, are reversed compared to Figs. 2e and 2f because of the flipping of the sign of topological charge of both the spatiotemporal vortex and the spatial vortex. The average OAM per photon of the wave packet is calculated to be $2.5\hbar$ pointing to the direction of 23.8 degrees with respect to the *y* axis in the *y-z'* plane.

The experimental demonstration shown in Fig. 3 is the intersection of two spatiotemporal vortices of topological charge -1 and 1 respectively with a spatial vortex of topological charge -1 in the same wave packet. Figures 3a to 3g are seven slices of interference fringe patterns. The locations of these slices are marked in Fig. 3h. Figures 3a to 3c show the interference patterns as the reference pulse slices through the intersection of the first spatiotemporal vortex with the spatial vortex. The upper half of peripheral fringes first bend to the left, then bend to the right after the π phase shift location. The first spatiotemporal vortex is therefore of topological charge -1 as also supported by the vortex tunnel twisting direction shown in Fig. 3h. Figure 3d shows a slice between the two intersections where the spatial vortex plays a dominant role over the spatiotemporal vortices. Consequently, the most noticeable feature is a forklike pattern pointing upwards at the center. Figures 3e to 3g display the interference patterns around the second intersection. The second spatiotemporal vortex has a topological charge 1 that results in a different bending sequence shown in Figs. 3e to 3g and an opposite twisting direction shown in Fig. 3h. This is the first experimental demonstration of the generation of ultrafast wave packet embedded with multiple optical vortices with individually tunable orientation. The average OAM per photon of the wave packet in the head is estimated to be 3.4$\hbar$ pointing to the direction of 17.3 degrees with respect to the *y* axis in the *y-z'* plane; the average OAM of the wave packet in the tail is estimated to be to 6.2$\hbar$ pointing to the direction of 9.2 degrees with respect to the *-y* axis in the *y-z'* plane. In this particular example, the two vortices with different OAM orientation are temporally separated by ~1 ps. With shorter pulses, more OAM vortices with temporal separation less than 100 fs packed within one wave packet can be expected.

**Conclusions**

In summary, we report the first experimental demonstration of the intersection of spatiotemporal vortices and spatial vortices. The twisting of vortex tunnels reveals the three-dimensional characteristics of optical vortices. The spatial vortex generates a longitudinal optical OAM whereas the spatiotemporal vortex provides a transverse OAM. In the intersection region where the two types of vortices interact most intensely, the vortex tunnels of both types of vortices are intertwined and twisted indicating a complex three-dimensional spiral phase structure. The associated OAM with the three-dimensional spiral phase structure is tilted with respect to the optical axis. The OAM carried by the wave packet consists of a longitudinal component and a transverse component that can be controlled individually through their corresponding topological charges. The average three-dimensional OAM per photon contained in the wave packet remain unchanged after propagation in free space. The tilted angle of the OAM provides a new degree of freedom that could possibly for used in optical tweezing, spin-orbit angular momentum coupling, and quantum communications. The first theoretical prediction on tilted OAM originates from an ordinary optical vortex with a fast-moving observer or source [11]. The experimental demonstration in this article provides a nonrelativistic way to generate tilted OAM. The platform, therefore, could potentially be utilized for simulating relativistic effects as well.

**Supplementary data**

Supplementary data are available at NSR online.

**Funding**

This work was supported by the National Natural Science Foundation of China (92050202, 61805142, 61875245), Shanghai Science and Technology Committee (19060502500), Shanghai Natural Science Foundation (20ZR1437600), and Wuhan Science and Technology Bureau (2020010601012169).

**Author contributions**

A.C. and C.W. proposed the original idea and performed all experiments and theoretical analysis. J.C. contributed in developing the measurement method. Q.Z. guided the theoretical analysis and supervised the project. All authors contributed to writing the manuscript.

*Conflict of interest statement.* None declared.

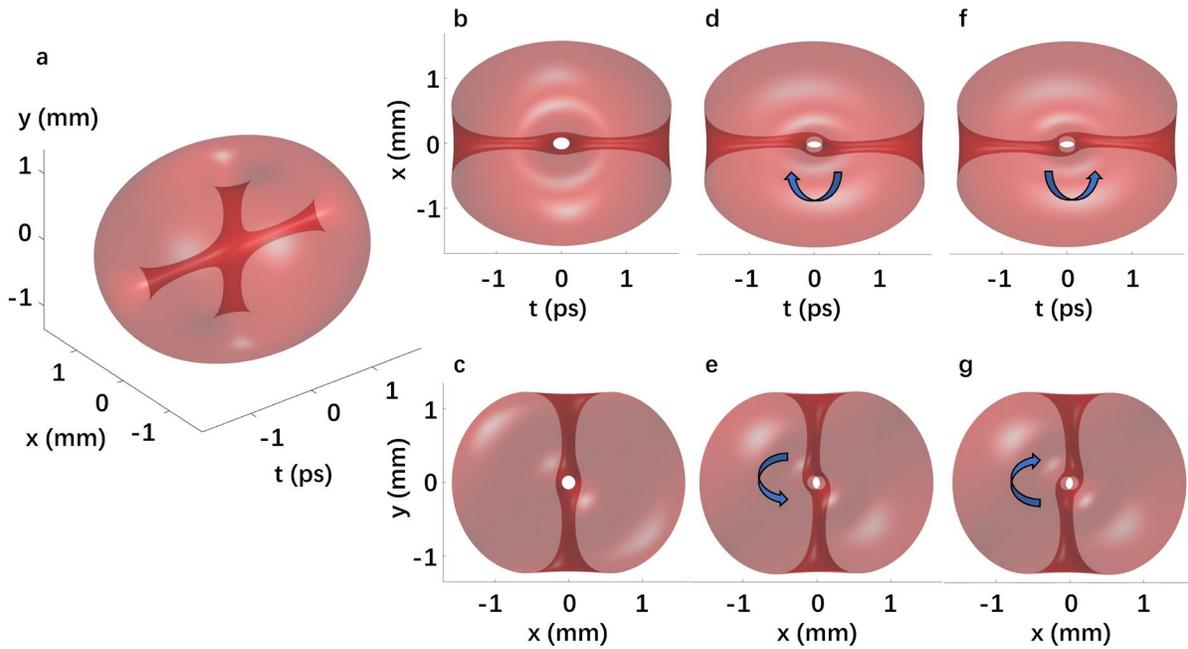

**Fig. 1. Simulated three-dimensional iso-intensity profiles of a wave packet with collided spatiotemporal and spatial vortices.** Normalized symmetric intensity distribution and unitary topological charge are assumed. The isovalue is set to 0.02 and 96.8% energy of the wave packet is contained in the isosurface. Fresnel diffraction theory is used for propagation of 12 cm. (a-c) A generated wave packet with collided vortices before propagation. (d,e) A generated wave packet with collided vortices after propagation. The topological charges of the spatiotemporal vortex and spatial vortex are both +1. (f,g) A generated wave packet with collided vortices after propagation. The topological charges of the spatiotemporal vortex and spatial vortex are both -1.

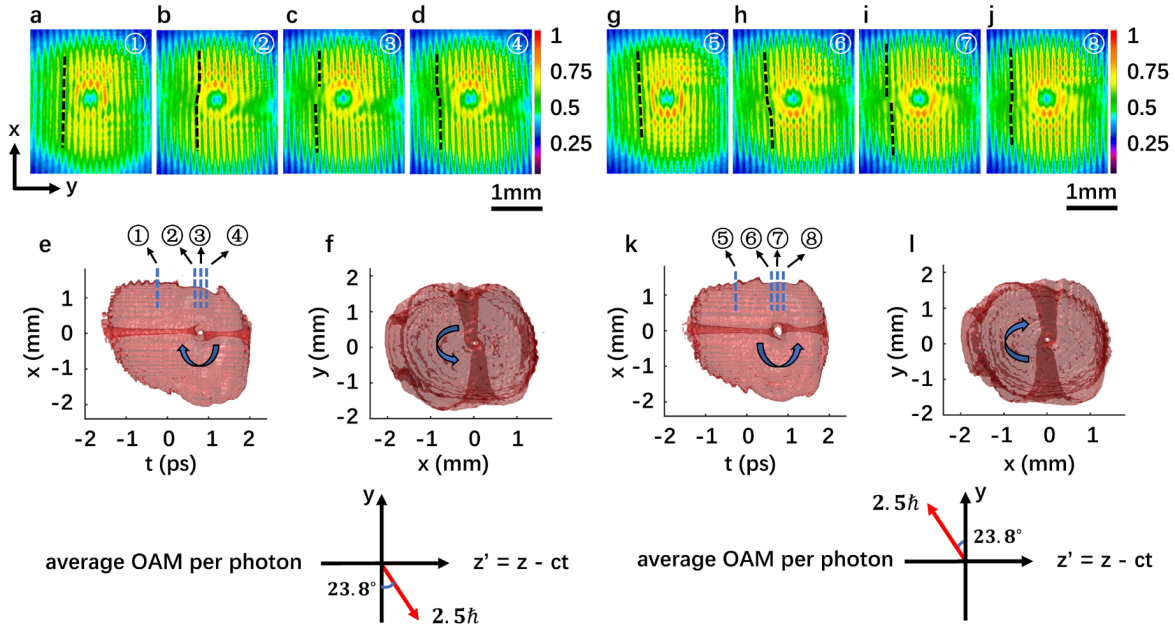

**Fig. 2. The intersection of a spatiotemporal vortex with a spatial vortex.** (a-d) The topological charge of the spatiotemporal vortex and the spatial vortex are both 1. Interference fringe patterns of the reference pulse with the chirped wave packet at various positions. Black dashed lines mark the bending directions of fringes. (e,f) Three-dimensional intensity reconstruction of the chirped wave packet showing the intersection of a spatiotemporal vortex and a spatial vortex from different views. 91.0% energy of the wave packet is contained in the isosurface. Temporal separation between slice 1,2,3 and 4 are 1ps, 83fs, 83fs, respectively. (g-j) The topological charge of the spatiotemporal vortex and the spatial vortex are both -1. Interference fringe patterns of the reference pulse with the chirped wave packet at various positions. (k,l) Three-dimensional intensity reconstruction of the chirped wave packet showing the intersection of a spatiotemporal vortex and a spatial vortex from different views. 89.8% energy of the wave packet is contained in the isosurface. Temporal separation between slice 1,2,3 and 4 are 1ps, 83fs, 83fs, respectively.

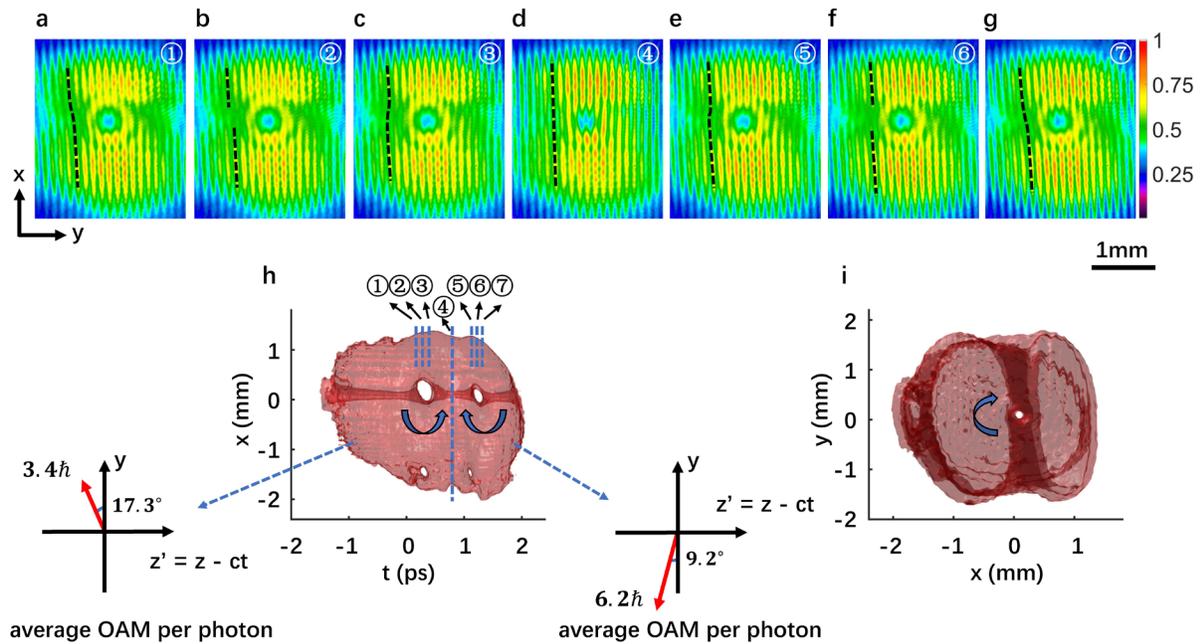

**Fig. 3. The intersection of two spatiotemporal vortices of topological charge -1 and 1 respectively with a spatial vortex of topological charge -1.** (a-g) Interference fringe patterns of the reference pulse with the chirped wave packet at various positions. Black dashed lines mark the bending directions of fringes. (h,i) Three-dimensional intensity reconstruction of the chirped wave packet showing the intersection of two spatiotemporal vortices and a spatial vortex from different views. 88.0% energy of the wave packet is contained in the isosurface. Temporal separation between slice 1-7 are 83fs, 83fs, 383fs, 383fs, 83fs, 83fs, respectively.